\documentclass[sigconf]{acmart}

\usepackage{graphicx}
\usepackage{float}
\AtBeginDocument{%
  }

\setcopyright{rightsretained}
\copyrightyear{2024}
\acmYear{2024}
\acmDOI{10.1145/3675094.3677545}

\acmConference[UbiComp Companion '24]{Companion of the 2024 ACM International Joint Conference on Pervasive and Ubiquitous Computing }{October 5--9, 2024}{Melbourne, VIC, Australia}
\acmBooktitle{Companion of the 2024 ACM International Joint Conference on Pervasive and Ubiquitous Computing (UbiComp Companion '24), October 5--9, 2024, Melbourne, VIC, Australia}
\acmISBN{979-8-4007-1058-2/24/10}

\begin{document}

\title[Enabling On-Device LLMs Personalization with Smartphone Sensing]{Enabling On-Device LLMs Personalization with \\Smartphone Sensing}

\author{Shiquan Zhang}
\affiliation{%
  \institution{The University of Melbourne}
  \city{Melbourne}
  \country{Australia}}
\email{shiquan.zhang@student.unimelb.edu.au}

\author{Ying Ma}
\affiliation{%
  \institution{The University of Melbourne}
  \city{Melbourne}
  \country{Australia}
}
  \email{ying.ma1@student.unimelb.edu.au}

\author{Le Fang}
\affiliation{%
 \institution{The University of Melbourne}
 \city{Melbourne}
  \country{Australia}
    }
  \email{le.fang1@student.unimelb.edu.au}

\author{Hong Jia}
\affiliation{%
 \institution{The University of Melbourne}
 \city{Melbourne}
  \country{Australia}
    }
\email{hong.jia@unimelb.edu.au}

\author{Simon D'Alfonso}
\affiliation{%
 \institution{The University of Melbourne}
 \city{Melbourne}
  \country{Australia}
    }
      \email{dalfonso@unimelb.edu.au}

\author{Vassilis Kostakos}
\affiliation{%
 \institution{The University of Melbourne}
 \city{Melbourne}
  \country{Australia}
  }
\email{vassilis.kostakos@unimelb.edu.au}

\renewcommand{\shortauthors}{Shiquan Zhang, Ying Ma, Le Fang, Hong Jia, Simon D'Alfonso, \& Vassilis Kostakos}

\begin{teaserfigure}
    \centering
    \includegraphics[width=\textwidth]{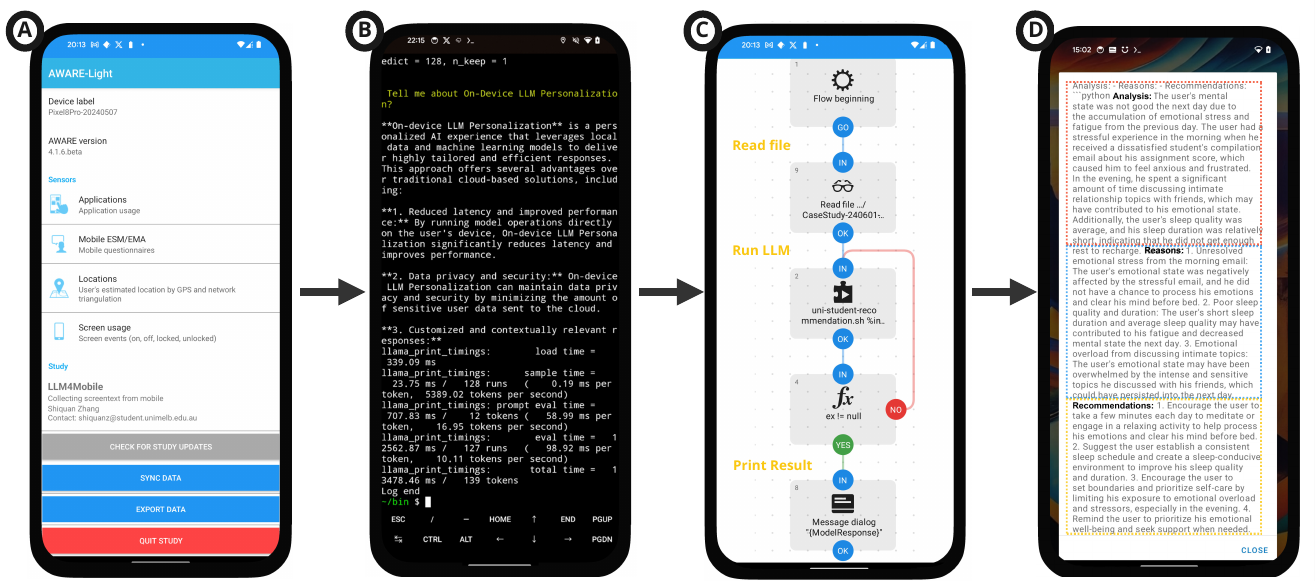}
    \caption{Overview of our work. (A) Sensing data collection from Aware-Light. (B) Implementation of LLM on smartphones. (C) Automatic workflow and trigger in Automate. (D) Personalized analysis and recommendations using LLM.}
    \Description{descrpiton of the figure but not show these lines of words in fig.}
    \label{fig:banner}
\end{teaserfigure}

\begin{abstract}
This demo presents a novel end-to-end framework that combines on-device large language models (LLMs) with smartphone sensing technologies to achieve context-aware and personalized services. The framework addresses critical limitations of current personalization solutions via cloud LLMs, such as privacy concerns, latency and cost, and limited personal information. To achieve this, we innovatively proposed deploying LLMs on smartphones with multimodal sensor data through context-aware sensing and customized prompt engineering, ensuring privacy and enhancing personalization performance. A case study involving a university student demonstrated the capability of the framework to provide tailored recommendations. In addition, we show that the framework achieves the best trade-off in privacy, performance, latency, cost, battery and energy consumption between on-device and cloud LLMs. To the best of our knowledge, this is the first framework to provide on-device LLMs personalization with smartphone sensing. Future work will incorporate more diverse sensor data and involve extensive user studies to enhance personalization. Our proposed framework has the potential to substantially improve user experiences across domains including healthcare, productivity, and entertainment.
\end{abstract}


\received{7 Jun 2024}
\received[revised]{1 July 2024}
\received[accepted]{1 July 2024}

\maketitle

\section{Introduction}
\label{introduction}
The emergence of large language models (LLMs), exemplified by ChatGPT~\cite{R1-1}, has revolutionized human-machine interaction and impacted people from all walks of life by leveraging vast amounts of data and sophisticated algorithms to provide users with flexibility and personalized content~\cite{R2}. Despite their impressive capabilities in understanding, reasoning, and generation, current LLMs face significant limitations, primarily related to privacy and security, since most contemporary LLMs (e.g., ChatGPT, Claude~\cite{R1-2}, and Gemini~\cite{R1-3}), primarily operate in cloud environments which require users to upload their personal data to the cloud, potentially leading to the leakage of sensitive personal information. 

Secondly, there are latency and cost issues to consider. Latency can significantly impact user experience, especially when the network is unstable or facing high request volumes and major outrage~\cite{R18}. In critical situations like healthcare monitoring, where real-time data analysis and responses are essential, such latency is unacceptable as it could potentially compromise patient care and safety. When considering cost, cloud LLM services are expensive, preventing their extensive usage. For example, calling APIs from servers can cost around 75 USD per million tokens, as seen with Claude 3 Opus~\cite{R19-2}.

Lastly, the inability of cloud LLMs to adapt to real-time contextual data from users poses a significant challenge to personalization. Existing LLMs require pre-collected datasets, hindering their usage in personalization scenarios that rely heavily on streaming data. When users request tasks beyond their boundaries, generic LLMs often fail to provide accurate results. Given these issues, it is reasonable to consider smartphones as the ideal platform for sensing human activities and delivering personalized services, as LLMs are not yet extensively deployed on smartphones for this purpose.

In this demo, we aim to address aforementioned challenges by leveraging on-device LLMs combined with smartphone sensing technologies to enable personalized and context-aware services. The objective is to develop an end-to-end framework for on-device personalization with personal multimodal information and customized prompt engineering to meet individual user needs. By overcoming privacy concerns, latency and cost, and limited personal sensor data, our approach can provide more secure, context-aware, and efficient personalized services that directly proceed on users' devices, paving the way for broader applications and improved user experiences in various domains such as healthcare, productivity, and entertainment. To verify the proposed framework, we present a case study exploring a day in the life of a university student and discuss the comparison between on-device and cloud LLMs. To the best of our knowledge, this is the first framework to provide on-device LLMs personalization with smartphone sensing.
\section{Related Work}
\label{related_work}
This section summarizes related studies on current developments and challenges regarding on-device LLMs, smartphone sensing and personalization.

\subsection{On-Device LLMs} \label{On-Device}
There is a growing trend towards creating smaller models for deployment on edge devices such as smartphones and wearable devices~\cite{R4, R5, R3, R6}. On-device LLMs refer to LLMs running locally on devices rather than in the cloud, which mitigates the concerns regarding privacy and latency, as the processing occurs locally without connecting to the Internet. Compared to centralized cloud LLMs, which can be plagued by latency and bandwidth issues, on-device LLMs provides faster, more reliable, and more efficient processing, leading to quicker, safer and privacy-preserved decision-making. However, deploying LLMs on edge devices presents challenges due to limited computational resources. Open-source models such as Llama-3-8B~\cite{R7}, Phi-3-mini~\cite{R6}, and Gemma-2B~\cite{R8} have been introduced and explored for deployment on devices such as PCs and smartphones. Efforts to deploy lightweight LLMs to edge devices are still at an infant stage and applications based on these models have not been comprehensively studied~\cite{R9, R10, R11, R12, R13, R20, R21}. In comparison, we are the first framework to provide on-device LLMs personalization on smartphones.

Although on-device smaller LLMs have advantages over larger LLMs, they still struggle to meet individual requirements without the inclusion of sufficient contextual knowledge from additional sources.  

\subsection{Smartphone Sensing and Personalization}
With the growing prevalence of smartphones and advancements in their sensing capabilities, smartphones have become a natural platform for understanding the interaction between users, machines, and environments. Smartphone sensing offers unique advantages in terms of cost-effectiveness, user acceptance, and the ability to capture context-aware, fine-grained, and continuous data streams~\cite{R14}. Meanwhile, personalization has become a research focus which aims to tailor services, content, and user experiences based on individual preferences, behaviours, and contexts~\cite{R22}. Therefore, smartphones, being inherently personal devices, are ideally suited to facilitate personalization by providing a rich source of user-specific and real-time data. 

These works~\cite{R15, R16, R17, R23}, however, either focus on collecting in-depth sensor data without exploring the potential of LLMs or are missing a comprehensive understanding of multi-modal sensors. In comparison, we propose to investigate LLM personalization with extensive multi-modal sensor data.
\section{Methods}
\label{method}
This part involves (1) collecting sensor data using AWARE-Light in Section \ref{awarelight}, (2) deploying LLMs on smartphones in Section \ref{llmonandroid}, (3) applying prompt engineering in Section \ref{promptengineering}, and integrating them into (4) an end-to-end pipeline framework in Section \ref{pframework}.

\subsection{Sensing Data from AWARE-Light}\label{awarelight}
AWARE-Light\footnote{\url{https://www.aware-light.org/}}~\cite{R15} is an open-source Android software for conducting smartphone sensing studies, which allows users to collect rich sensor data and deploy scheduled questionnaires from both hardware and software on smartphones. It has extensive sensors\footnote{\url{https://www.aware-light.org/sensors/}} such as geolocation, accelerometer, experience sampling method (ESM), keyboard, communications, app usage and screen text. In this demo, we utilized screentext and ESM sensors to collect data, where the screentext sensor ~\cite{R16} can capture all screen text on smartphones, and mobile ESM sensor can periodically collect questionnaire data. Notably, to accommodate on-device scenarios, we specifically extended an export function to export sensor data to local Android files.

\subsection{LLMs on Android} \label{llmonandroid}
We built a working environment to run LLMs on smartphones. Termux\footnote{\url{https://termux.dev/en/}}, an Android terminal emulator and Linux environment app, was installed on a Google Pixel 8 Pro with 12GB DRAM. Then, a machine learning compilation engine, a high-performance universal deployment solution that allows native deployment of large language models with compiler acceleration, was set up from llama.cpp\footnote{\url{https://github.com/ggerganov/llama.cpp}}, which enables LLMs to run locally with acceleration in Android. Open-source lightweight models, Llama-3-8B ~\cite{R8}, were downloaded through Hugging Face\footnote{\url{https://huggingface.co/}} and deployed on the phone. Lastly, LLMs can run locally with a chat interface like ChatGPT in Fig~\ref{fig:banner}.B.

\begin{figure}[h]
    \centering
    \includegraphics[width=0.35\textwidth]{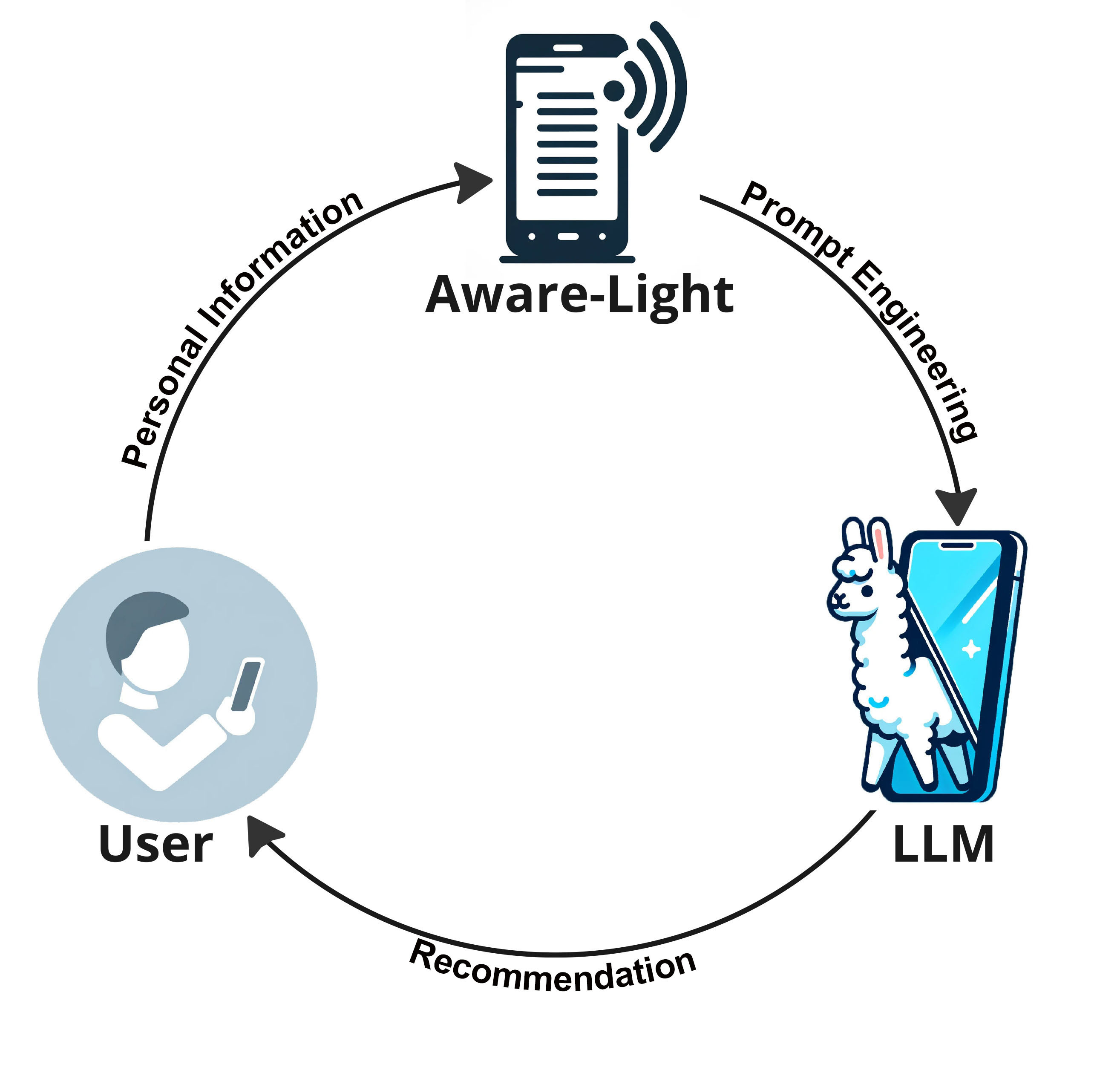}
    \caption{Pipeline of this end-to-end framework. Sensing data collected through Aware-Light from users is processed on the phone with LLMs using prompt engineering to generate personalized responses for users.}
    \label{fig:pipeline}
\end{figure}

\subsection{Prompt Engineering Design} \label{promptengineering}
As defined in Equations~\ref{eq:promptengineering_1} and ~\ref{eq:promptengineering_2} , a structured prompt template ~\cite{P24} was designed to give LLMs sufficient contextual information. \textit{Instruction} provides the model with explicit guidance on what task needs to be performed, which sets the objective and helps narrow down the model's response to the desired action. \textit{Context} gives the model the background information necessary to understand the task and produce relevant responses, which is significant to deliver personalization. In this situation, \textit{Context} involves $C_{user}$\textit{}, personal information, and \textit{$C_{domain}$}, a specific area of knowledge, and \textit{$C_{sensing}$}, sensing data from AWARE-Light. \textit{Question} specifies what exactly the user wants to know or the information that the user seeks from the model. \textit{Output Format} defines the format in which user wants the information to be presented.
\begin{equation}
Prompt= Instruction + Context + Question + Output Format
\label{eq:promptengineering_1}
\end{equation}
\begin{equation}
Context= C_{user} + C_{domain} + C_{sensing} 
\label{eq:promptengineering_2}
\end{equation}

\begin{figure*}[h]
    \centering
    \includegraphics[width=\linewidth]{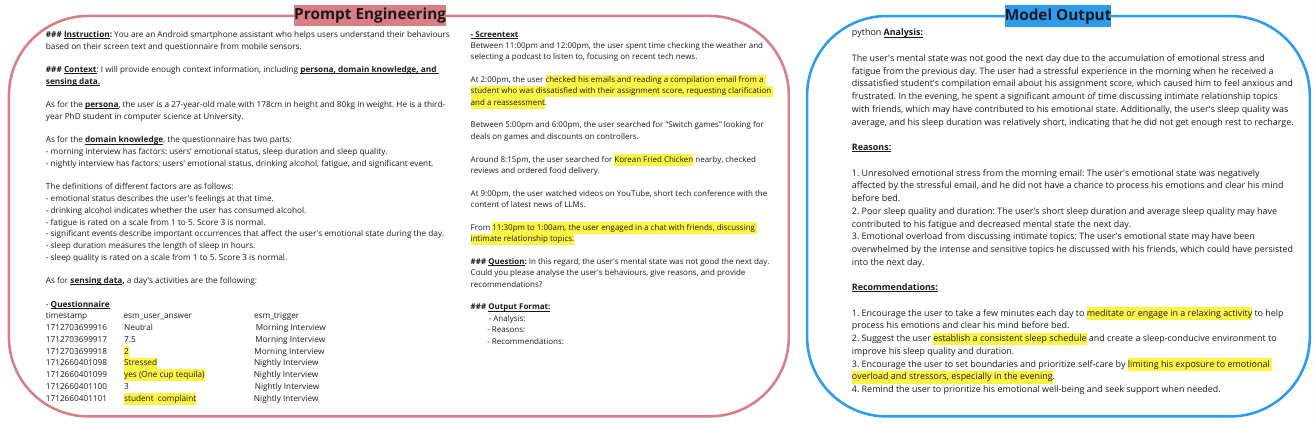}
    \caption{Prompt Engineering Design and Model Output.}
    \label{fig:inout}
\end{figure*}

\subsection{An End-to-End Pipeline Framework} \label{pframework}
Fig~\ref{fig:pipeline} illustrates an end-to-end pipeline framework, where sensing data will be collected through AWARE-Light from users, LLMs with prompt engineering will be processed on the phone and finally generate personalized responses to users. Automate\footnote{\url{https://llamalab.com/automate/}}, an Android automation tool, is used to trigger the pipeline and visualize the generated responses from LLMs.
\section{Experimental Setup and Evaluation}
\label{Experimental_Setup_and_Evaluation}
In this part, a case study in Session~\ref{CaseStudy} was explored under this framework and the comparison of on-device and cloud LLMs were discussed.

\subsection{Case Study} \label{CaseStudy}
A day of smartphone activity from a university student was captured through screentext and mobile ESM sensors from AWARE-Light. Regarding screentext, fine-grained text data were collected. Considering the verbose input, local Llama-3-8B was utilized to summarize the collected screentext. Regarding ESM questionnaires, morning and nightly questions were delivered at 9 AM and 9 PM, respectively,  covering users' emotional status, sleep duration, sleep quality, fatigue, alcohol consumption, and significant event factors.

A mental state problem was raised to test our framework. More details can be found on prompt design and model responses in Fig~\ref{fig:inout}. Preliminary analysis can be observed from sensor data. Via the ESM questionnaire, the user expressed stress, suffered from a ``student complaint'' and ``drunk alcohol on nightly interview'' while had a ``bad sleep'' even with ``average sleep hours last night'' on morning interview. As for screentext, an email about ``a student complain about assignment score from tutoring'', ``unhealthy dinner'', and ``chatting intimate relationship in late night'' may impact his mental state.

The local LLM provided insightful analysis and explanation based on the captured information, highlighting unresolved emotional stress from a complaint email, poor sleep quality, and intimate nighttime conversations. Personalized analysis and suggestions were made, ``meditate or engage in a relaxing activity before bed'', ``establish a consistent sleep schedule'', and ``limiting exposure to emotional overload and stressors'', demonstrating the potential for personalized services such as  counseling, coaching, and co-piloting.

However, local models still have limitations. Specifically, we occasionally observed that these models exhibit hallucinations, such as generating unrelated content and reversing concepts, as well as ignorance, like failing to consider certain information. Session~\ref{futurework} will discuss potential approaches to mitigate these problems.

\subsection{On-Device and Cloud LLMs Comparison}
Both cloud and on-device LLMs can deliver personalized services in the case study. Compared to cloud LLMs, our work focusing on on-device LLMs prioritizes \textbf{privacy}, as all processing is performed on the smartphone. 
In terms of \textbf{performance}, cloud LLMs can leverage state-of-the-art models, while on-device LLMs may be inferior in accuracy and have a less up-to-date knowledge base. However, we envision future pocket LLMs will be more lightweight, robust and accurate. 
Regarding \textbf{latency}, cloud LLMs can experience higher latency due to potential network instability or high request volumes and major outrage~\cite{R18}, affecting real-time performance. In comparison, our framework can provide faster and more reliable responses. 
In terms of \textbf{cost}, on-device LLMs are free, while cloud LLMs require calling APIs from servers such as ChatGPT-4o~\cite{R19-1}, Claude 3 Opus~\cite{R19-2}, Gemini 1.5 Pro~\cite{R19-3}, which can cost from 5 to 75 USD per million tokens. 
When considering \textbf{battery consumption}, on-device LLMs directly impact the device's battery life. Our tests indicated approximately 16.5\% RAM usage and 3\% battery drain in 5 minutes on a Google Pixel 8 Pro during our case study on on-device model inference, highlighting one of the future research factors for on-device LLMs.
Finally, when examining \textbf{energy consumption}, cloud LLMs consume considerable energy due to their extensive computational requirements. An LLM consumes 0.1 J per token for every billion parameters~\cite{R3}. For example, the cloud LLM ChatGPT, with 175B parameters, consumes approximately 17.5 J per token, resulting in substantial energy usage. On-device LLMs, while consuming less energy overall, still contribute to energy usage, albeit to a lesser extent. For instance, a 7B-parameter model consumes 0.7 J per token. Moreover, the substantial energy consumption of cloud LLMs has significant environmental implications, with ongoing energy usage contributing to increased carbon dioxide emissions and exacerbating environmental problems. As such, enabling on-device LLMs could make significant contributions for ecological AI.
\section{Demonstration}
\label{Demonstration}
We plan to show a live demo on Google Pixel 8 Pro smartphone, offering an immersive and interactive experience. Users can begin by exploring the AWARE-Light app, observing how it collects various types of sensor data, such as GPS, screentext and ESM. This will provide insights into the foundational data that powers our personalized recommendations.
Next, users can engage with on-device LLMs directly on smartphones, where they can create their own prompts or modify pre-built prompts with their sensor data to run LLMs.
Lastly, users can view and iterate personalized recommendations based on their interactions and the collected sensor data. 
This live demo will highlight the seamless integration of data collection, prompting, and personalized recommendation generation, showcasing the robustness and practicality of our end-to-end framework on smartphones.
\section{Conclusion \& Future Work} 
\label{Conclusion_futurework}
We presented a novel end-to-end framework that showcases on-device LLMs to provide more secure, context-aware, and efficient personalized services based on smartphone sensing. Additionally, we conducted a case study to test this framework and deliver personalized recommendations, and made comparison between on-device and cloud LLMs. Our initial experiments demonstrated the great potential of this framework for personalized services. In future work, considering some occasional errors, more contextual information, domain knowledge, and user-specific models, such as fine-tuning models for university scenarios, should be considered. Moreover, we plan to integrate more diverse sensor data and conduct large-scale user studies, allowing us to refine this framework and determine how to deliver more personalized services effectively.




\bibliographystyle{sections/ACM-Reference-Format}
\bibliography{sections/main}

\end{document}